\documentclass[12pt]{iopart}
\usepackage{iopams}
\usepackage{amssymb}
\usepackage{graphicx}
\usepackage{color}
\usepackage[tight,TABTOPCAP]{subfigure}
\usepackage{epstopdf}
\begin{document}
\makeatletter

\newbox\slashbox \setbox\slashbox=\hbox{$/$}
\newbox\Slashbox \setbox\Slashbox=\hbox{$/$}
\def\pFMslash#1{\setbox\@tempboxa=\hbox{$#1$}
  \@tempdima=0.5\wd\slashbox \advance\@tempdima 0.5\wd\@tempboxa
  \copy\slashbox \kern-\@tempdima \box\@tempboxa}
\def\pFMSlash#1{\setbox\@tempboxa=\hbox{$#1$}
  \@tempdima=0.5\wd\Slashbox \advance\@tempdima 0.5\wd\@tempboxa
  \copy\Slashbox \kern-\@tempdima \box\@tempboxa}
\def\FMslash{\protect\pFMslash}
\def\FMSlash{\protect\pFMSlash}
\def\miss#1{\ifmmode{/\mkern-11mu #1}\else{${/\mkern-11mu #1}$}\fi}
\makeatother


\title[Bounding the $Z^\prime tc$ coupling...]{Bounding the $Z^\prime tc$ coupling from $D^{0}-\overline{D^{0}}$ mixing and single top production at the ILC}
\author{J. I. Aranda$^{(a)}$, F. Ram\'\i rez-Zavaleta$^{(a)}$, J. J. Toscano$^{(b)}$, and E. S. Tututi$^{(a)}$}
\address{$^{(a)}$Facultad de Ciencias F\'\i sico Matem\' aticas,
Universidad Michoacana de San Nicol\'as de
Hidalgo, Avenida Francisco J. M\'ujica S/N, 58060, Morelia, Michoac\'an, M\' exico. \\
$^{(b)}$Facultad de Ciencias F\'{\i}sico Matem\'aticas,
Benem\'erita Universidad Aut\'onoma de Puebla, Apartado Postal
1152, Puebla, Puebla, M\'exico.}

\begin{abstract}
In the present work the $Z^\prime tc$ coupling is bounded by using the current experimental data on the $D^{0}-\overline{D^{0}}$ meson mixing system. It is found that the strength associated to this coupling is less than $5.75\times 10^{-2}$. The single top production through the $e^+e^-\to Z^\prime\to tc$ process at the $Z^\prime$ boson resonance is studied and we found that around $10^7$ $tc$ events will be expected at the International Linear Collider. For the $Z^\prime\to tc$ decay, we predict a branching ratio of $10^{-2}$.
\end{abstract}

\pacs{14.70.Pw, 13.38.Dg, 14.65.Ha}

\maketitle

\section{Introduction}
Many extensions of the Standard Model (SM) predict the existence of an extra $U^{\prime}(1)$ gauge symmetry group and its associated $Z^{\prime}$ boson which has been object of extensive phenomenological studies~\cite{langacker1,leike,langacker-rmp}. This  boson can induce flavor changing neutral currents (FCNC) at the tree level through $Z^{\prime}q_{i}q_{j}$ couplings where $q_{i}$ and $q_{j}$ are up or down-type quarks. Due to the large  top quark mass, it is commonly believed that the top quark physics could show up a window for research on new physics effects beyond the SM~\cite{Larios}, FCNC mediated by a $Z^{\prime}$ neutral gauge boson in particular. The FCNC couplings of the top quark that are susceptible of being observed have been examined within the context of extended models~\cite{langacker2,arhrib,cakir,tosca,valencia1,val-a,giladp,val-a1,aranda1,aranda2,frank}. The simplest extended model that predicts an extra neutral gauge boson, identified as $Z^{\prime}$, is based on the $SU_L(2)\times U_Y(1)\times U^{\prime}(1)$ extended electroweak gauge group, which after spontaneous symmetry breaking (SSB) generates a mixing between the $Z$ and $Z^{\prime}$ neutral gauge bosons~\cite{langacker2}. Nevertheless, from electroweak precision data, it has been established that the  corresponding  mixing angle  is strongly suppressed~\cite{langacker-rmp,langacker2}. The FCNC couplings have been studied within the context of non-universality, where it is assumed that the strength in the flavor diagonal couplings differs in the third family~\cite{arhrib,cakir,valencia1,val-a,giladp,val-a1}. A numerical simulation study through the $pp\to tZ^\prime$ processes considering a signal yield of 10 events for an integrated luminosity of 100 $\mathrm{fb}^{-1}$ was performed at LHC scenario~\cite{val-a1}, in which the $Z^{\prime} tc$ and $Z^{\prime} tu$ couplings strength were found to be  of orders of $10^{-2}$ and $10^{-3}$, respectively. On the other hand, the single top production through the $pp\to tc$ and $e^+e^-\to tc$ reactions, was studied at the LHC and ILC, respectively~\cite{arhrib}. In this work the strength of $Z^{\prime}cu$ coupling was bounded from $D^0-\overline{D^0}$ mixing. In the same context of non-universality, in Ref.~\cite{cakir} the  top quark FCNC couplings  through $Z^{\prime}$ exchange at the LHC and CLIC are studied to calculate the single top  FCNC production cross section in the $pp\to tc X$ and $e^+e^- \to Z^{\prime}\to  tc$ processes, respectively.

The flavor-violating parameters must fulfill FCNC experimental constraints. Some models such as the left-right symmetric model or the $SO(10)$  grand unification model  predict that the down-type quark transitions are strongly suppressed, while the up-type quark  transitions can be as large as the $U_{ts}$ CKM element. Particularly, the strength of the $Z^{\prime} tu$ coupling is comparable to the $U_{ub}$ element~\cite{val-a}. Some studies on $t \to c$ flavor-changing transitions assume that they could manifest themselves as a rare top quark decays. These type of transitions do not produce constraints on $Z^\prime tc$ couplings since the $Z^\prime$ boson is required to be heavier than the top quark, provided the virtual $Z^\prime$ effects are not taken into account. In contrast, if we focus on the $Z^\prime$ virtual effects we may analyze the impact of the FCNC through the single top quark production.

In this work, the $Z^{\prime} tu_i$ ($u_i=u, c$) FCNC couplings are studied. We use the mass difference  $\Delta M_D$ of the $D^{0}-\overline{D^{0}}$ mixing observed by the Babar \cite{babar} and Belle \cite{bel} collaborations  to bound the strength of these couplings. To accomplish this task, it is necessary to take into account contributions arising from diagrams at the tree  and one-loop  level. As we shall see, the dominant contribution for the $Z^{\prime}tc$ coupling comes from the tree level diagram, implying a stronger constraint than the resulting one from the one-loop calculation. For the $Z^\prime \to q_iq_j$ process, with $q_i,q_j=u,c,t$, the general renormalizable FCNC couplings coming from the model based on the $SU_L(2)\times U_Y(1)\times U^{\prime}(1)$ gauge group are used. We restrict ourselves to the aforementioned couplings, since our objective is to compute the $tc$ production rate mediated by a $Z^\prime$ gauge boson at the ILC.  Instead of  using the $x$  parameter of non-universality~\cite{arhrib,cakir} to express the strength of the FCNC couplings in terms of it,  we  treat the strength  of these couplings  as a parameters to be determined in the spirit of a model-independent approach. In this way, it leads  to results  such that the  only free parameter is the $Z^{\prime}$ boson mass; this allows us to express the strength of the coupling as a function of the $Z^{\prime}$ mass boson. Then  we calculate the cross section associated with the $e^+e^-\to tc$ reaction mediated by a $Z^\prime$ gauge boson at the ILC scenario and estimate the branching ratio for the $Z^\prime \to tc$ decay process. An outline of this work is as follows. In section~\ref{FG}, we briefly present the theoretical frame of work for the model used. In section~\ref{FG1}, we calculate the $Z^{\prime} q_iq_j$ couplings from the $D^{0}-\overline{D^{0}}$ meson mixing system. In section~\ref{Collider}, we compute the $e^{+}e^{-} \to tc$ FCNC process at the tree level by using the constraint on the $Z^{\prime} tc$ coupling and estimate the branching ratio for $Z^\prime\to tc$ decay. In section~\ref{final}, we discuss our results.

\section{The $Z^{\prime} tu_{i}$ couplings}\label{FG}

The FCNC Lagrangian contained in the simplest extended model based on the $SU_C(3)\times SU_L(2)\times U_Y(1)\times U^{\prime}(1)$ electroweak gauge group~\cite{durkin, langacker3} is given by
\begin{equation}\label{general}
\mathcal{L}_{NC}=-eJ_{EM}^\mu A_{\mu}-g_1J_1^\mu Z_{\mu,1}-g_2 J_{2}^{\mu} Z_{{\mu},2},
\end{equation}
where $e$ is the electromagnetic coupling, $J^\mu_{EM}$ is the electromagnetic neutral current, $g_1$ is the gauge coupling of the SM, $J^\mu_1$ is the weak neutral current of the SM, $g_2$ is the gauge coupling of the $U^{\prime}(1)$ group and $J_2^\mu$ represents the new weak neutral current given as
\begin{equation}\label{fotW}
J_{2}^{\mu}=\sum_{i,j} \overline{\psi^\prime_i} \, \, \gamma^{\mu} (\epsilon_{L{ij}}^{\psi}\, P_L+\epsilon_{R{ij}}^{\psi}\, P_R)\, \psi^\prime_j ,
\end{equation}
where $\epsilon_{L,R{ij}}^{\psi}$ are the chiral couplings of $Z_2$ with $i,j$ running over all leptons and quarks,  $\psi^\prime_i$  represents a fermion in the gauge interaction basis, and $P_{L,R}=\frac{1}{2}(1\pm \gamma_{5})$ are the chiral projectors. Since the interaction between the bosons $Z_{1}$ and $Z_2$ is too weak to be considered, we suppose no mixing between them, consequently their mass eigenstates are $Z^0$ and $Z^\prime$, respectively~\cite{langacker-rmp,langacker2}. Since  we are interested in the flavor-violating $tc$ production rate, we consider the $\epsilon_{L,R{ij}}^{u}$ matrix for the up quark sector. Some models  assume this matrix as flavor diagonal and  non-universal, where the simplest treatment is the one in which the third element $\epsilon_{tt}^{u}$ is different~\cite{arhrib,cakir,langa}. In this work we also assume a $\epsilon_{L,R}^{u}$ matrix with the same properties. The FCNC couplings $\epsilon_{R,L}^{u}$, which are in the gauge eigenstates basis, are transformed into the mass eigenstates ones by diagonalizing the mass matrix in the Yukawa sector~\cite{durkin,langacker3}. Therefore, the FCNC couplings in the mass eigenstates basis ($\Omega_L, \Omega_R$) can be read off as
\begin{eqnarray}\label{fotW2}
\Omega_{{L}ij}&=\,g_2\,(V_{L}\, \epsilon_{L}^{u} \, V_{L}^{\dagger})_{ij},\nonumber\\
\Omega_{{R}ij}&=\,g_2\,(V_{R}\, \epsilon_{R}^{u} \, V_{R}^{\dagger})_{ij},
\end{eqnarray}
where $V_{L,R}$ are  the unitary matrices that diagonalize the mass matrix in the Yukawa sector of the SM. The up quark current given by equation~(\ref{fotW}) is transformed into the mass eigenstates basis as
\begin{eqnarray}
\label{fotW3} J^{\mu}&=\sum_{i,j} \overline{u}_i \, \, \gamma^{\mu} (\Omega_{{L}ij}\, P_L+\Omega_{{R}ij}\, P_R)\, u_j,
\end{eqnarray}
where $u_i=u,c,t$.
This current allows us to study the FCNC transitions $Z^{\prime} tu_{i}$, which are given by the strength of the $\Omega_{{L,R}\,tu_{i}}$ matrices.

\section{Bounding the  $Z^{\prime} tc$ couplings from $
D^{0}-\overline{D^{0}}$}\label{FG1}

Our main concern in this section is to bound the strength of the $Z^\prime q_iq_j$ $(i\neq j)$ coupling. To carry out this task we resort to the experimental result for the $D^{0}-\overline{D^{0}}$ meson-mixing system~\cite{babar,bel}. According to equations~(\ref{general}) and (\ref{fotW3}), the part of the Lagrangian containing the relevant information is
\begin{eqnarray}
\label{diagramas}
\mathcal{L}_{NC}^{Z^\prime q_iq_j}&=-\big[\overline{u}\,
\gamma^{\mu} (\Omega_{{L}uc} \, P_L+\Omega_{{R}uc} \,
P_R)\, c+\overline{c}\, \gamma^{\mu} (\Omega_{{L}cu} \,
P_L+\Omega_{{R}cu} \, P_R) \, u
\nonumber\\
&+\overline{u}\,
\gamma^{\mu} (\Omega_{{L}ut} \, P_L+\Omega_{{R}ut} \,
P_R)\, t+\overline{t}\,
\gamma^{\mu} (\Omega_{{L}tu} \, P_L+\Omega_{{R}tu} \,
P_R) \, u
\nonumber\\
&+\overline{c}\, \gamma^{\mu} (\Omega_{{L}ct} \,
P_L+\Omega_{{R}ct} \, P_R) \, t+\overline{t}\, \gamma^{\mu} (\Omega_{{L}tc} \,
P_L+\Omega_{{R}tc} \, P_R) \, c\big]Z^\prime_\mu.
\end{eqnarray}

It is easy to see, from the unitary property of the $V_{L,R}$ matrices, that
\begin{equation}
|\Omega_{uc}| \approx |\Omega_{ut}\Omega_{ct}|,
\label{omega-relation}
\end{equation}
provided that $\epsilon_{tt}\ll 1$ and $|V_{{L,R}\,q_{i}q_{j}}|\ll 1$ for $i\neq j$. Since we focus only on  bounding the strength of the coupling $Z^\prime q_iq_j$, regardless of the possible CP-violating effects, we assume for simplicity that the different $\Omega$'s are real, $\Omega_{{L,R}\,q_iq_j}=\Omega_{{L,R}\,q_jq_i}$ and $\Omega_{L\,q_iq_j}=\Omega_{R\,q_iq_j}\equiv \Omega_{q_iq_j}$.  To constraint the $|\Omega_{tc}|$ parameter, it is necessary to take into account the short-distance effects which are included in the tree-level and box diagrams shown in figure~\ref{Figuras1}.  Because  the amplitudes are dominated by the $Z^\prime$ gauge boson and top quark    masses, we may neglect the external momenta (heavy mass limit).
In this approximation, the tree-level amplitude can be written as
\begin{equation}
{\cal M}_{\rm tree}=-\frac{i\Omega^{2}_{uc}}{m_{Z^\prime}^2}\,\,\overline{u}\gamma^{\alpha}c\,\,\overline{u}\gamma_{\alpha}c,
\label{tree-level-amp}
\end{equation}
where $m_{Z^\prime}$ is the mass of the $Z^{\prime}$ gauge boson.
The $\cal{M}_{\rm tree}$ amplitude is related to a four-quark effective interaction given by the effective Lagrangian:
\begin{equation}
{\cal L}^{\rm tree}_{eff}=-\frac{\Omega^{2}_{uc}}{4m^{2}_{Z^\prime}}\left(Q_1+2 Q_2+Q_6\right),
\label{lag-tree}
\end{equation}
where a $1/4$ factor has been introduced to compensate two Wick contractions. The various $Q_i$ in equation~(\ref{lag-tree}) are dimension 6 effective operators used in the literature~\cite{pakvasa,p-pakvasa}.

Analogously, the one-loop level amplitude is given by:
\begin{figure}[ht]
\centering
\subfigure[]{\includegraphics[scale=0.6]{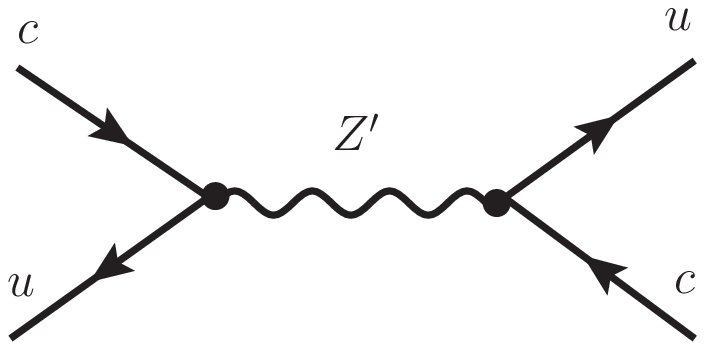}}\qquad\qquad
\subfigure[]{\includegraphics[scale=0.6]{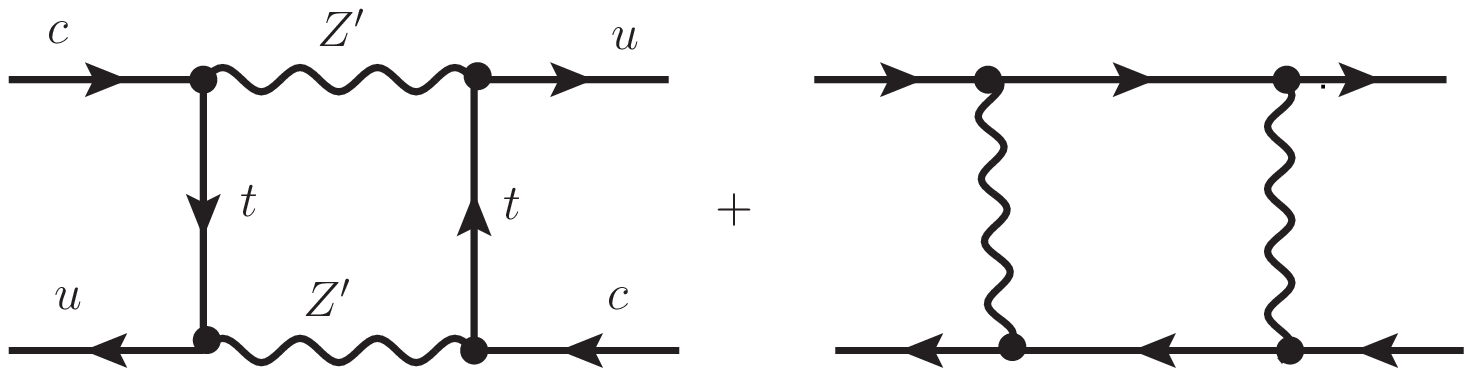}}
\caption{(a) Tree diagram (b) Box diagrams for $ D^{0}-\overline{D^{0}}$ mixing.}
\label{Figuras1}
\end{figure}
\begin{eqnarray}
\label{diagramas2} \cal{M}_{\rm box}&=\,2\, \Omega^2_{tu}\, \Omega^2_{tc}\,
\int \frac{d^{\,4}k}{(2\,\pi)^{4}}\,  \frac{[\overline{u}\,
\gamma^{\lambda}\, (\pFMSlash{k}+m_{t})\, \gamma^{\nu}\, c]\,
[\overline{u}\, \gamma_{\nu}\, (\pFMSlash{k}+m_{t})\,
\gamma_{\lambda}\, c]}{(k^{2}-m_{t}^{2})^{2}\,
(k^{2}-m_{Z^\prime}^{2})^{2}},
\end{eqnarray}
where $m_t$ denotes the mass of the top quark.
After some algebra we arrive at the following result:
\begin{eqnarray}
\label{diagramas3} \cal{M}_{\rm box}&=-\frac{\it{i}}{16\pi^{2}}\, \,
 \frac{\Omega_{tu}^{2}\, \Omega_{tc}^{2}}{m_{t}^{2}}\,  \Big [f(x)\,
 \overline{u}\, \gamma^{\lambda} \gamma^{\alpha} \gamma^{\nu}\, c\, \overline{u}\, \gamma_{\nu} \gamma_{\alpha} \gamma_{\lambda}\, c\nonumber\\
 &+g(x)\,\overline{u}\, \gamma^{\lambda} \gamma^{\nu}\, c\, \overline{u}\, \gamma_{\nu}  \gamma_{\lambda}\, c\Big ],
\end{eqnarray}
where $f(x)$ and $g(x)$ are loop functions given as
\begin{eqnarray}
\label{diagramas4} f(x)&=\frac{1}{2}\,
 \frac{1}{(1-x)^{3}}\,  [1-x^{2}+2\,x\,\log x],\\
 g(x)&=\frac{2}{(1-x)^{3}}\,  [2(1-x)+(1+x)\,\log x].
\end{eqnarray}
with $x=m_{Z^\prime}^{2}/m_{t}^{2}$.

The $\cal{M}_{\rm box}$ amplitude is related to a four-quark effective interaction given by the effective Lagrangian:
\begin{eqnarray}
\label{diagramas5} \mathcal{L}_{eff}^{\rm box}&=-
 \frac{\Omega_{tu}^{2}\, \Omega_{tc}^{2}}{64 \pi^{2} m_{t}^{2}}\,  \big [f(x)\,
 (4Q_{1}+32 Q_{2}+4Q_{6})\nonumber\\
 &+g(x)\,(8 Q_{3}+4 Q_{4}+Q_{5}+4 Q_{7}+Q_{8})\big ],
\end{eqnarray}
 again a $1/4$ factor has been introduced to compensate two Wick contractions.
 On the other hand, the mass difference $\Delta M_D$ provided by the $D^{0}-\overline{D^{0}}$ meson-mixing system is given by

\begin{eqnarray}
\label{diagramas6} \Delta M_{D}&=
 \frac{1}{M_{D}}\,  Re \langle \overline{D^{0}}| \mathcal{H}_{eff}=-\mathcal{L}_{eff}|D^{0} \rangle,
\end{eqnarray}
where $\mathcal{H}_{eff}$ is the effective Hamiltonian and $M_D$ is the $D^0$ meson mass. Here, the effective Lagrangian is obtained by adding    the tree-level  and box contributions: $\mathcal{L}_{eff}={\cal L}^{\rm tree}_{eff}+{\cal L}^{\rm box}_{eff}$. Therefore, by using the modified vacuum saturation approximation~\cite{pakvasa} we have:
\begin{eqnarray}
\Delta M_{D}&=\frac{\Omega^{2}_{uc}}{4m^{2}_{Z^\prime}}\bigg[\langle Q_1\rangle+2 \langle Q_2\rangle+\langle Q_6\rangle+\frac{x}{16\pi^2}\Big(f(x)
 (4\langle Q_{1}\rangle+32\langle Q_{2}\rangle\nonumber\\
 &+4 \langle Q_{6}\rangle)+ g(x)\,(8 \langle Q_{3}\rangle+4 \langle Q_{4}\rangle+Q_{5}+4 \langle Q_{7}\rangle+\langle Q_{8}\rangle)\Big)\bigg]
\nonumber\\
&=\frac{1}{12}\frac{\Omega^{2}_{uc}}{m^{2}_{Z^\prime}}f^{2}_{D}M_DB_D\left[1+\frac{x}{8\pi^2}\big(32 f(x)-5g(x)\big)\right],
\label{deltam}
\end{eqnarray}
where we have used the relation given by equation~(\ref{omega-relation}), $B_D$ is the bag model parameter and $f_D$ represents the $D^0$ meson decay constant. We can see from  equations~(\ref{lag-tree}), (\ref{diagramas5}) and (\ref{deltam}) that the main contribution to $\Delta M_D$ comes from the tree-level amplitude while the contribution coming from the box amplitude  is of approximately  17\%-19\% in the range of $800\,\, \mathrm{GeV}\leq m_{Z^\prime}\leq 3000\,\, \mathrm{GeV}$.

In order to compute the strength of the couplings we take $B_D\sim1$, $f_D=222.6$ MeV~\cite{cleo} and $M_D=1.8646$ GeV~\cite{pdg}.
Considering that $\Delta M_D$  does not exceed the experimental uncertainty~\cite{babar,bel,heavyflavor}, the bound  is
\begin{equation}
|\Omega_{uc}|<\frac{3.6\times 10^{-7}m_{Z^\prime}\,{\rm GeV}^{-1}}{\sqrt{1+\frac{x}{8\pi^2}\left(32 f(x)-5g(x)\right)}},
\label{tree-contr}
\end{equation}
\begin{figure}[ht]
\centering
\includegraphics[scale=0.9]{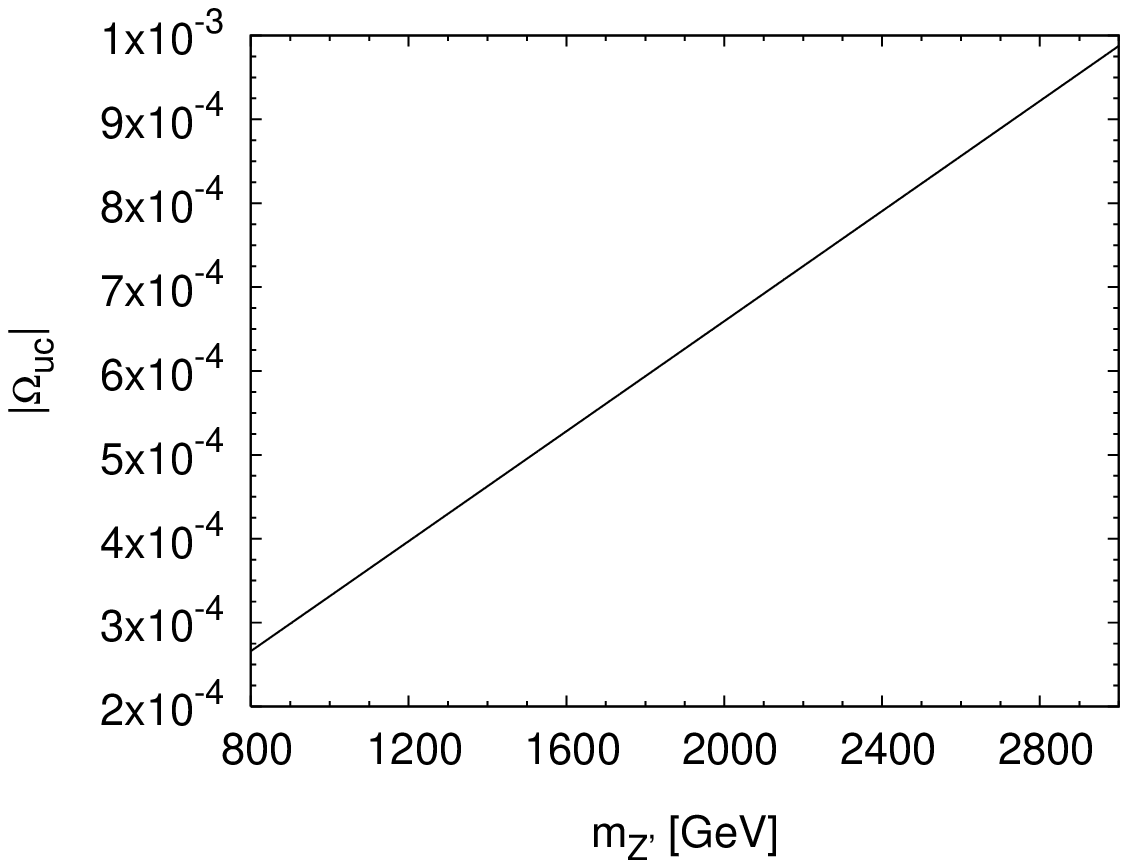}
\caption{Behavior of $|\Omega_{uc}|$ coupling  as a function of $Z^{\prime}$ boson mass.}
\label{Figuras2}
\end{figure}
In figure~\ref{Figuras2}, we depict the behavior of maximum of $|\Omega_{uc}|$  as a function of the $Z^\prime$ boson mass. It can be observed that the growth is not bigger than one order of magnitude in a broad range of the mass.

Since we are interested in bounding the $\Omega_{tc}$ coupling, we take the relation previously found: $|\Omega_{uc}|\approx |\Omega_{tc} \Omega_{tu}|$. For $m_{Z^\prime}=1$ TeV, we obtain a bound $|\Omega_{tc}\Omega_{tu}|<3.31 \times 10^{-4}$, moreover, if we assume that $\Omega_{tc}=10\,\Omega_{tu}$, as it occurs for the absolute values of $U_{ts}, U_{td}$ elements in the CKM matrix~\cite{pdg}, we found that  $|\Omega_{tc}|<5.75\times 10^{-2}$  and $|\Omega_{tu}|<5.75\times 10^{-3}$, which are of the same order of magnitude than those obtained in~\cite{arhrib,val-a}. Let us emphasize that our bound on $Z^\prime tc$ coupling was calculated by using experimental results on the $D^{0}-\overline{D^{0}}$ meson-mixing system. We can also mention that the  value  $\Omega_{tc}=10\,\Omega_{tu}$ taken to bound the $Z^\prime tc$ coupling strength is equivalent to the value assumed of few tenths for the  $x$  parameter of non-universality in other approaches~\cite{arhrib,cakir}; however, our assumption is inspired by  physical results taken from the CKM matrix.

\section{The process $e^+e^-\to Z^\prime \to tc$ at ILC collider}\label{Collider}

In previous works~\cite{arhrib,cakir,tosca}, the flavor-diagonal $Z^\prime f\bar f$ ($f$ being any SM fermion) couplings have been calculated, each having different strength depending on the model used. To make predictions on the process  $e^+e^-\to Z^\prime \to tc$, from those couplings, here we use the upper and the lower values provided by the sequential $Z$ and $E_6$\footnote{Although in Ref.~\cite{arhrib,cakir} the  strength parameter with the biggest values is for the $Z_{\psi}$ model which is contained in a larger group $E_6$, we simply have termed it as $E_6$ model.} models, respectively. The relevant parameters used to calculate the strength of the couplings are  the chiral charges $Q^{f}_{L,R}$ taken from~\cite{arhrib}. Rather than using the maximum and minimum values of the strength of these  chiral charges,  we find interesting, only  for comparison proposes, to  take  their average; the different values for the  charges are displayed in table~\ref{table}.  Additionally, we also use the strength of the $\Omega_{tc}\Omega_{tu}$ coupling previously determined to complete  the set of parameters needed to compute the total width for the $Z^{\prime}$ boson decay.

\begin{table}
\caption{\label{table}The strength of the flavor-diagonal $Z^\prime f\bar f$ couplings.}
\footnotesize\rm
\begin{tabular*}{\textwidth}{@{}l*{15}{@{\extracolsep{0pt plus12pt}}l}}
\br
& Sequential $Z$ & $E_6$ & Average \\
\mr
$Q_L^u$ & $0.3456$ &
      $\frac{1}{\sqrt{24}}$ & $0.2749$ \\

$Q_R^u$ & $-0.1544$  &
      $\frac{-1}{\sqrt{24}}$ & $-0.1793$ \\

$Q_L^d$ & $-0.4228$ &
      $\frac{1}{\sqrt{24}}$ & $-0.1093$ \\

$Q_R^d$ & $0.0772$ &
      $\frac{-1}{\sqrt{24}}$ & $-0.0635$ \\

$Q_L^e$ & $-0.2684$ &
      $\frac{1}{\sqrt{24}}$ &  $-0.0321$ \\

$Q_R^e$ & $0.2316$ &
      $\frac{-1}{\sqrt{24}}$ & $0.0137$ \\

$Q_L^\nu$ & $0.5$ &
      $\frac{1}{\sqrt{24}}$ &  $0.3521$ \\
\br
\end{tabular*}
\end{table}
We resort to the Breit-Wigner resonant cross section, which is
\begin{equation}
\sigma(e^+e^-\to Z^\prime \to tc)=\frac{12\,\pi\,m_{Z^\prime}^2}{s}\frac{\Gamma(Z^\prime\to e^+e^-)\,\Gamma(Z^\prime\to tc)}{(s-m_{Z^\prime}^2)^2+m_{Z^\prime}^2\Gamma_{Z^\prime}^2}
\label{cross}
\end{equation}
where $\sqrt{s}$ is the center of mass energy and $\Gamma_{Z^\prime}$ is the $Z^\prime$ total decay width. On account of the total $Z^\prime$ width we include the total possible flavor-diagonal and a flavor-nondiagonal decay modes, namely: $\nu_e\bar\nu_e$, $\nu_\mu\bar\nu_\mu$, $\nu_\tau\bar\nu_\tau$, $e^+e^-$, $\mu^+\mu^-$, $\tau^+\tau^-$, $u\bar u$, $c\bar c$, $t\bar t$, $d\bar d$, $s\bar s$, $b\bar b$, $\bar uc+u\bar c$, and $\bar tc+t\bar c$. For the width of the $Z^\prime\to e^+e^-$ process we have employed the results given in~\cite{arhrib}. For the decay width $\Gamma(Z^\prime\to tc)$, we obtain
\begin{equation}
\Gamma(Z^\prime\to tc)=\frac{\left(2 m_{Z^\prime}^4-m_{t}^4-m_{Z^\prime}^2 m_{t}^2\right) {\Omega^2_{tc}}}{12\,\pi\, m_{Z^\prime}^3 }.
\end{equation}

\begin{figure}[ht]
\centering
\includegraphics[scale=.9]{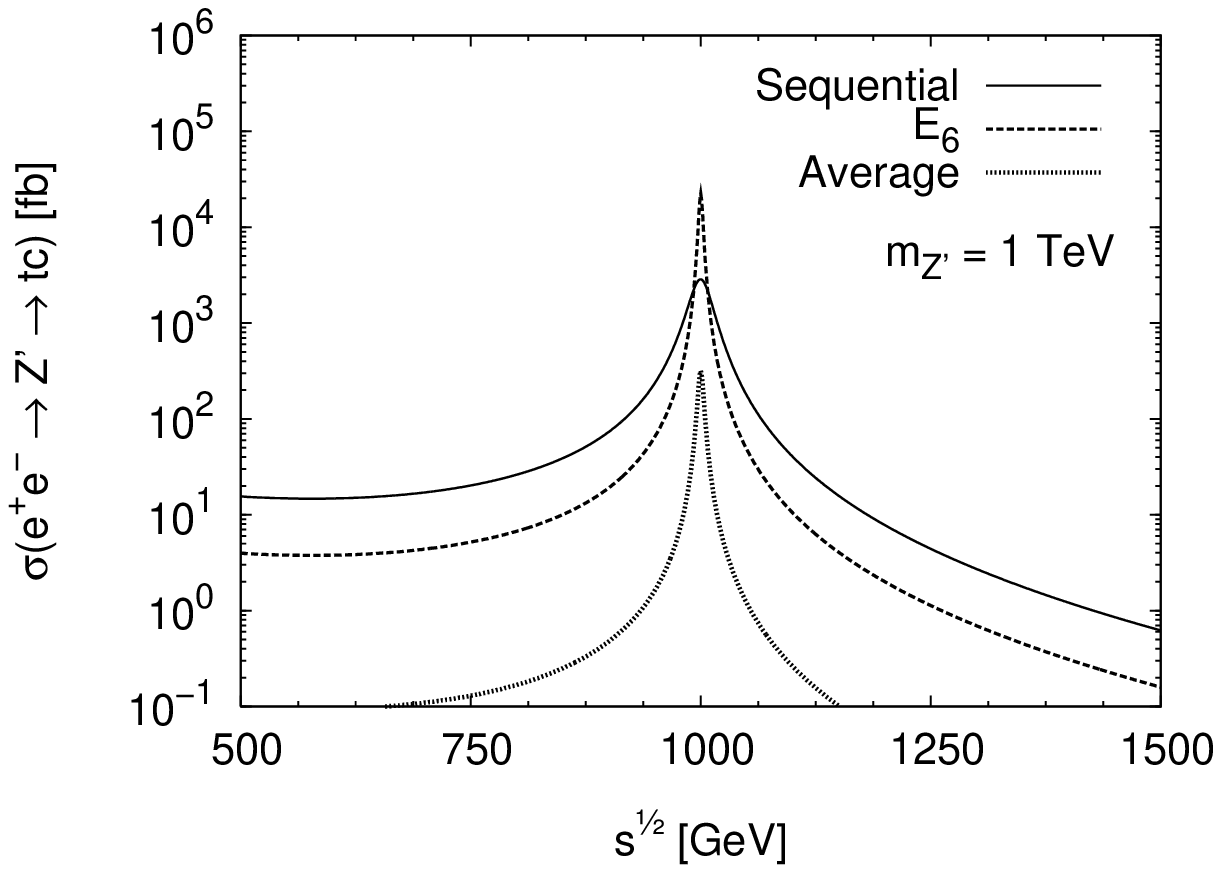}
\caption{\label{CSS}Cross section for $e^+e^-\to Z^\prime \to tc$ process as a function of $\sqrt{s}$ for $m_{Z^\prime}=1$ TeV.}
\end{figure}

\begin{figure}[ht]
\centering
\includegraphics[scale=.9]{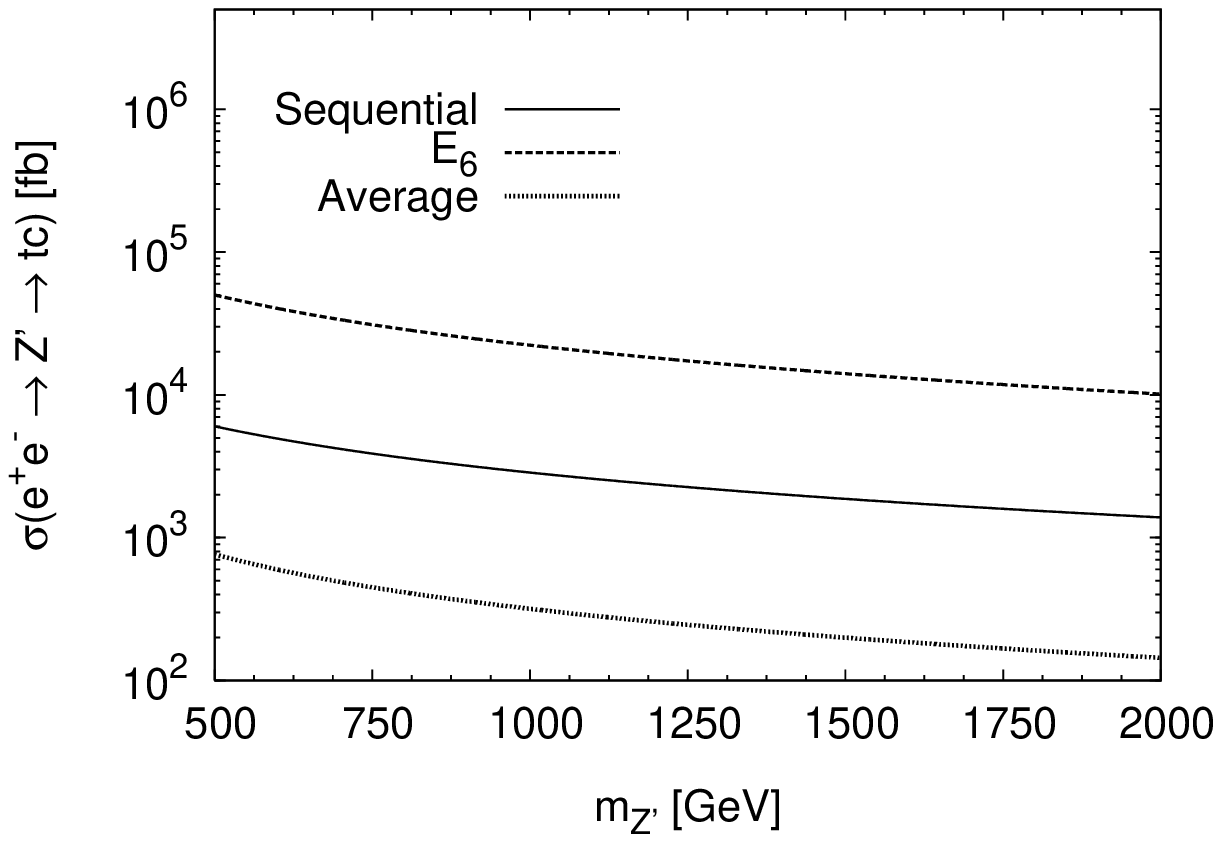}
\caption{\label{CSR}Cross section for $e^+e^-\to Z^\prime \to tc$ process at the resonant values as a function of $m_{Z^\prime}$.}
\end{figure}

\begin{figure}
\centering
\includegraphics[scale=.9]{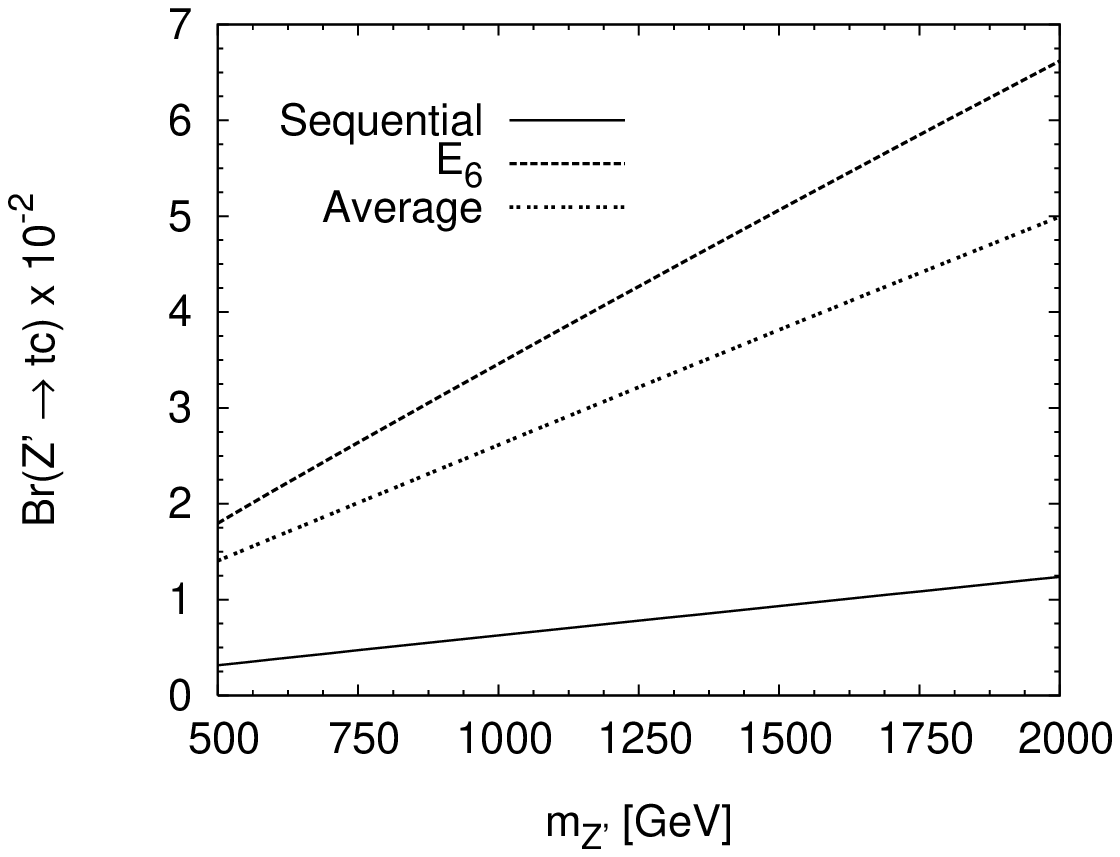}
\caption{\label{BR}The branching ratio for $Z^\prime \to tc$ decay.}
\end{figure}

In figure~\ref{CSS} we show the Breit-Wigner cross section for the $tc$ flavor violation production as a function of the center-of-mass energy. For computing the cross section, we set the same value of the $\Omega_{tc}$ parameter for both analyzed models as a function of the $Z^\prime$ boson mass. It is expected that the ILC will operate at a luminosity of $500$ $\mathrm{fb}^{-1}$ in the first years of running~\cite{ILC}. In this context, from our results shown in figure~\ref{CSR}, we can predict around $10^7$ events just at the resonance for the $E_6$ model, while for the sequential $Z$ model it is expected to obtain around $10^6$ events and for the average of the two models, it is expected around $10^5$ events. According to our results, it is more feasible to observe flavor violation in the $E_6$ model, which is corroborated with the $Z^\prime \to tc$ decay as it can be appreciated from figure~\ref{BR}. We obtain that the associated branching ratio is of the order of $10^{-2}$.

It is useful to compare our results on single top quark production mediated by  a $Z^{\prime}$ gauge boson with others previously obtained from different models and approaches.  In the context of the LHC, for the single top quark production with flavor violation mediated by a $Z^\prime$, a previous exhaustive work has developed that the capability of the LHC to detect a $tc$ FCNC effects by measuring the production of $t\bar c+\bar tc$ pairs is much less as compared to the ILC approach~\cite{arhrib}. So we leave aside this context and concentrate on linear colliders scenarios. First, we compare our results for the $e^+ e^-\to\gamma\gamma\to tc$ cross section at the energy $\sqrt{s}= 500$ GeV (see figure~\ref{CSS})  with that obtained in the SUSY model~\cite{yu}. The supersymmetric model predicts a cross section of the order of 1 fb, which agrees with our most conservative result; however, it is one order of magnitude smaller than our prediction for the sequential $Z$ model.

On the other hand,  within the context of the ILC, at the energy of the resonance, $\sqrt{s}\sim m_{Z^{\prime}}\sim 1$ TeV, our average gives around $10^5$ $t c$ events; this production is higher by four orders of magnitude than those derived from the SUSY models~\cite{frank}, which were calculated at  energies higher than $2m_t$. In addition, the production of around $10^4$ $t c$ events calculated at the resonance within the context of the Compact Linear Collider~\cite{arhrib,cakir} can be compared with our predictions which are higher by  one  and three orders of magnitude for the average    and  the $E_6$ model, respectively. The discrepancy in these results can be understood if we take into account that the values for the couplings used in~\cite{arhrib,cakir}, $(B^u)_{ct}=(V^{\dagger}\epsilon^u V)_{ct}\sim 10^{-2}$, are model dependent for both the sequential $Z$ and the $E_6$ models, while the  value for our coupling is model independent, $\Omega_{tc}\sim 10^{-1}$ and it was obtained from experimental restrictions.

Within the same context of the ILC, we may also contrast our result, with the obtained previously for  us~\cite{aranda2}, where the top quark  production is mediated by a Higgs boson. We found that around  $10^3$ $t c$ events will be produced for a Higgs mass of the order of top quark mass, which is two orders of magnitude less than    the average prediction, calculated at the resonance. Moreover, we have estimated the branching ratios for the $Z^\prime\to t c$ and $Z^\prime\to t u$ decays calculated at the resonance, which are of the order of $10^{-2}$ and $10^{-4}$, respectively. We can mention that these values are one order of magnitude less restrictive than the corresponding  branching ratios obtained in the 3-3-1 model~\cite{tosca}.

\section{Final remarks}\label{final}
In this work we have studied the possible flavor violation mediated by a $Z^\prime$ neutral gauge boson, which is predicted by several models beyond the SM. We have bounded the strength of the flavor-violating $Z^\prime tc$ coupling using the experimental results coming from the $D^{0}-\overline{D^{0}}$ meson-mixing system, where our constraint depends only  on the $Z^\prime$ boson mass.  For a $m_{Z^\prime}=1$ TeV, we have found that the flavor-violating parameter $|\Omega_{tc}|<5.75\times 10^{-2}$ is in agreement with similar predictions in previous works. We have calculated the cross section for the $e^+e^-\to Z^\prime \to tc$ process in the ILC scenario, where we found an estimation of around $10^7$ events for a luminosity of $500$ $\mathrm{fb}^{-1}$ in the context of $Z^\prime$ boson predicted by the $E_6$ model. According to our results, the $tc$ flavor violation effect mediated by a $Z^\prime$ boson from the $E_6$ model is more favorable of being observed than that predicted in the sequential model. This behavior is also repeated for the branching ratio of the $Z^\prime\to tc$ decay.

\section*{Acknowledgments}
This work has been partially supported by CONACYT and CIC-UMSNH. EST thanks to  FCFM-BUAP and the HEP group for their kind hospitality.\\

\end{document}